\documentclass{aastex631}

\submitjournal{ApJL}

\shorttitle{The Southern Galactic Pole Stream}
\shortauthors{Yang et al.}
\graphicspath{{./}{figures/}}

\begin{document}

\title{A New Cold Stream near the Southern Galactic Pole}

\correspondingauthor{Jing-Kun Zhao}
\email{zjk@bao.ac.cn}

\author[0000-0001-7609-1947]{Yong Yang}
\affiliation{CAS Key Laboratory of Optical Astronomy, National Astronomical Observatories, Chinese Academy of Sciences, Beijing 100101, People's Republic of China}
\affiliation{School of Astronomy and Space Science, University of Chinese Academy of Sciences, Beijing 100049, People's Republic of China},
\author[0000-0003-2868-8276]{Jing-Kun Zhao}
\affiliation{CAS Key Laboratory of Optical Astronomy, National Astronomical Observatories, Chinese Academy of Sciences, Beijing 100101, People's Republic of China},
\author[0000-0002-0642-5689]{Xiang-Xiang Xue}
\affiliation{CAS Key Laboratory of Optical Astronomy, National Astronomical Observatories, Chinese Academy of Sciences, Beijing 100101, People's Republic of China},
\author[0000-0002-5805-8112]{Xian-Hao Ye}
\affiliation{CAS Key Laboratory of Optical Astronomy, National Astronomical Observatories, Chinese Academy of Sciences, Beijing 100101, People's Republic of China}
\affiliation{School of Astronomy and Space Science, University of Chinese Academy of Sciences, Beijing 100049, People's Republic of China},
\author[0000-0002-8980-945X]{Gang Zhao}
\affiliation{CAS Key Laboratory of Optical Astronomy, National Astronomical Observatories, Chinese Academy of Sciences, Beijing 100101, People's Republic of China}
\affiliation{School of Astronomy and Space Science, University of Chinese Academy of Sciences, Beijing 100049, People's Republic of China}



\begin{abstract}

We report the discovery of a cold stream near the southern Galactic pole (dubbed as SGP-S) detected in $Gaia$ Early Data Release 3. The stream is at a heliocentric distance of $\sim$ 9.5 kpc and spans nearly 58$\degr$ by 0.6$\degr$ on sky. The colour–magnitude diagram of SGP-S indicates an old and metal-poor (age $\sim$ 12 Gyr, [M/H] $\sim$ -2.0 dex) stellar population. The stream's surface brightness reaches an exceedingly low level of $\Sigma_G \simeq$ 36.2 mag arcsec$^{-2}$. Neither extant globular clusters nor other known streams are associated with SGP-S. 

\end{abstract}

\keywords{Milky Way Galaxy (1054) --- Milky Way stellar halo (1060) --- Tidal tails (1701) --- Globular star clusters (656)}


\section{Introduction} \label{sec:intro}

Increasing amount of data from revolutionary surveys are revealing that the Milky Way is full of substructures, either in the disk \citep[e.g.,][]{2009ApJ...692L.113Z,2018ApJ...868..105Z,2017ApJ...844..152L,2018A&A...619A..72R,2018Natur.561..360A,2021ApJ...922..105Y,2021ApJ...907L..16R,2021AJ....162..171Y,2021SCPMA..6439562Z}, or in the halo \citep[e.g.,][]{1994Natur.370..194I,2002ApJ...569..245N,2009ApJ...700L..61N,2010ApJ...714..229L,2018Natur.563...85H,2020ApJ...904...61Z}. Among those substructures, dynamically cold streams which are usually related to globular clusters (GCs) play an important role \citep[e.g.,][]{2006ApJ...639L..17G,2006ApJ...643L..17G,2009AJ....137.3378O,2012ApJ...760L...6B,2014MNRAS.442L..85K}. It has been proved that cold streams are powerful tools in constraining the Galactic potential and studying the formation history of the stellar halo \citep{2010ApJ...712..260K,2013MNRAS.436.2386L,2016ApJ...833...31B,2019MNRAS.486.2995M,2021ApJ...914..123I,2022MNRAS.513..853Y}.

In this letter, we report the discovery of a new cold stream near the southern Galactic pole which we designate SGP-S. The stream is exposed by weighting stars in color-magnitude diagram (CMD) and proper motions (PMs) simultaneously using $Gaia$ Early Data Release 3 (EDR3) \citep{2021A&A...649A...1G,2021A&A...649A...2L,2021A&A...649A...3R}. Section \ref{sec:detection} describes the detecting strategy and Section \ref{sec:stream} characterizes the stream. A conclusion is given in Section \ref{sec:summary}. 

\section{Weighting Stars} \label{sec:detection}

We first need to clarify that SGP-S was detected by chance during examining the existence of GC NGC 5824's leading tail. Specifically, \citet{2021ApJ...909L..26B} and \citet{2022ApJ...928...30L} pointed that Triangulum \citep{2012ApJ...760L...6B} and Turbio \citep{2018ApJ...862..114S} stream could be associated with NGC 5824. Motivated by this, we tried to search for other stream segments along its leading tail using a modified matched-filter technique from \citet{2019ApJ...884..174G}. The technique weighted stars using their color differences from the cluster’s locus in CMD. These weights are further scaled based on stars’ departures from PMs of the NGC 5824 model stream. By applying the method, we accidentally found the signature of SGP-S. 

After some experiments, we optimize choices of the filters and present details of the detection as follows. Stars from $Gaia$ EDR3 within a sky box of -20$\degr$ $<$ $\alpha$ $<$ 20$\degr$ and -90$\degr$ $<$ $\delta$ $<$ 40$\degr$ are retrieved. As \citet{2021A&A...649A...3R} pointed, for the source without measured $\nu_{\rm eff}$ (effective wavenumber) used in determining the $G$-band flux, a default $\nu_{\rm eff}$ is adopted and this will lead to a systematic effect in $G$-band photometry. In addition, \citet{2021A&A...649A...3R} introduced the corrected BP and RP flux excess factor $C^*$ to identify sources for which the $G$–band photometry and BP and RP photometry is not consistent. We correct the $G$ magnitude and calculate $C^*$ for our sample according to their Table 2 and 5. In order to ensure good astrometric and photometric solutions, only stars with \texttt{ruwe} $<$ 1.4 and $|C^*| < 3\sigma_{C^*}$ \citep[see Section 9.4 in][]{2021A&A...649A...3R} are retained. 

In CMD, we use a set of stellar tracks extracted from Padova database \citep{2012MNRAS.427..127B} at different distances as the filters. The isochrone grid explored here covers a metallicity range of -2.2 $\leq$ [M/H] $\leq$ -1.2 and an age range of 10 $\leq$ Age $\leq$ 13 with 0.1 dex and 1 Gyr spacing, along with a distance modulus (DM) varying from 14 to 17 with a step of 0.1 mag. Individual stars are assigned weights based on their color differences from a given isochrone filter, assuming a Gaussian error distribution:
\begin{equation}
	w_{\rm CMD} = \frac{1}{\sqrt{2\pi} \sigma_{color}} {\rm exp}
	\left[ -\frac{1}{2} \left(\frac{color - color_0}{\sigma_{color}} \right)^2  \right]  .
\end{equation}
Here $color$ and $\sigma_{color}$ denote $BP - RP$ and corresponding errors. $\sigma_{color}$ is simply calculated through $\sqrt{\sigma^2_{BP} + \sigma^2_{RP}}$ where $\sigma_{BP}$ and $\sigma_{RP}$ are obtained with a propagation of flux errors (see CDS website\footnote{\url{https://vizier.u-strasbg.fr/viz-bin/VizieR-n?-source=METAnot&catid=1350&notid=63&-out=text}.}). $color_0$ is determined by the isochrone at a given $G$ magnitude of a star. All stars have been extinction-corrected using the \citet{1998ApJ...500..525S} maps as re-calibrated by \citet{2011ApJ...737..103S} with $RV$ = 3.1, assuming $A_{G}/A_{V} = 0.83627$, $A_{BP}/A_{V} = 1.08337$, $A_{RP}/A_{V} = 0.63439$\footnote{These extinction ratios are listed on the Padova model site \url{http://stev.oapd.inaf.it/cgi-bin/cmd}.}. 

\begin{figure}
	\plotone{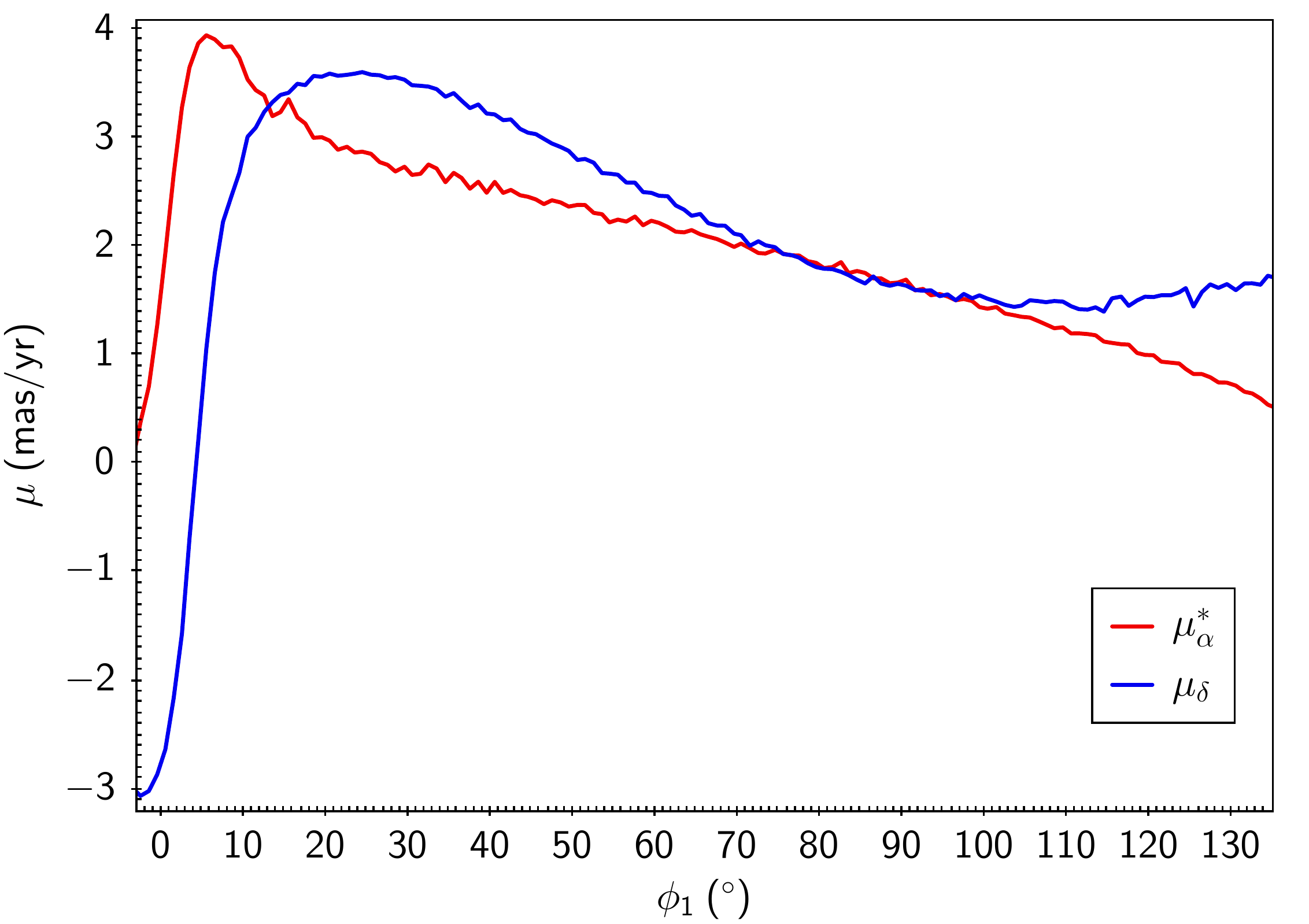}
	\caption{The PM filters used to weight stars with Equation \ref{equation2}. $\phi_1$ is the longitude in a rotated frame with a pole of ($\alpha$ = 270$\degr$, $\delta$ = 0$\degr$). The red line denotes $\mu^*_{\alpha}$ medians of NGC 5824 model stream particles in each $\phi_1$ bin with a bin width of 1$\degr$. Their $\mu_{\delta}$ medians are further added 1 mas/yr and shown with the blue line. 
	\label{fig:phi1_pm}}
\end{figure}

In terms of PMs, weights are computed as:
\begin{equation}
	w_{\rm PMs} = \frac{1}{2\pi \sigma_{\mu^*_{\alpha}}\sigma_{\mu_{\delta}}} {\rm exp}
	\left\lbrace -\frac{1}{2} \left[ \left( \frac{\mu^*_{\alpha} - \mu^*_{\alpha,0}}{\sigma_{\mu^*_{\alpha}}} \right)^2 +
	\left( \frac{\mu_{\delta} - \mu_{\delta,0}}{\sigma_{\mu_{\delta}}} \right)^2 \right] \right\rbrace .
	\label{equation2}
\end{equation}
Here $\mu^*_{\alpha}$, $\mu_{\delta}$, $\sigma_{\mu^*_{\alpha}}$ and $\sigma_{\mu_{\delta}}$ are measured PMs and corresponding errors. $\mu^*_{\alpha,0}$ and $\mu_{\delta,0}$ are the components of PMs predicted at each star's $\phi_1$ based on the run in Figure \ref{fig:phi1_pm}. For easier data processing and clearer presentation of results, we rotate celestial coordinates such that the point ($\alpha$ = 270$\degr$, $\delta$ = 0$\degr$) is the pole. Hence, $\phi_1$ here, which is the new longitude, corresponds to $\delta$ but starts at $\delta$ = -90$\degr$ and increases along $\alpha$ = 0$\degr$. The red and blue tracks come from the model stream of NGC 5824. The way of generating the model stream follows closely that of \citet{2022MNRAS.513..853Y} as applied to NGC 5466. Under a static Milky Way potential plus a moving Large Magellanic Cloud (LMC), GC NGC 5824 is initialized 2 Gyr ago and integrated forward from then on, releasing particles in both leading and trailing directions at Lagrange points \citep{2014MNRAS.445.3788G,2019MNRAS.487.2685E}. The resulting stream particles are divided into $\phi_1$ bins (bin width = 1$\degr$) and medians of PMs in each bin are calculated, by which the PM tracks of Figure \ref{fig:phi1_pm} are obtained. We further add 1 mas/yr to the whole $\mu_{\delta}$ because such a filter is a closer estimate to the $\mu_{\delta}$ trend of SGP-S and gives stronger signals (see below). 

Finally, stars weights are obtained by multiplying $w_{\rm CMD}$ and $w_{\rm PMs}$, and then summed in sky pixels to expose structures. 

\section{The SGP-S} \label{sec:stream}

\begin{figure*}
	\includegraphics[width=\linewidth]{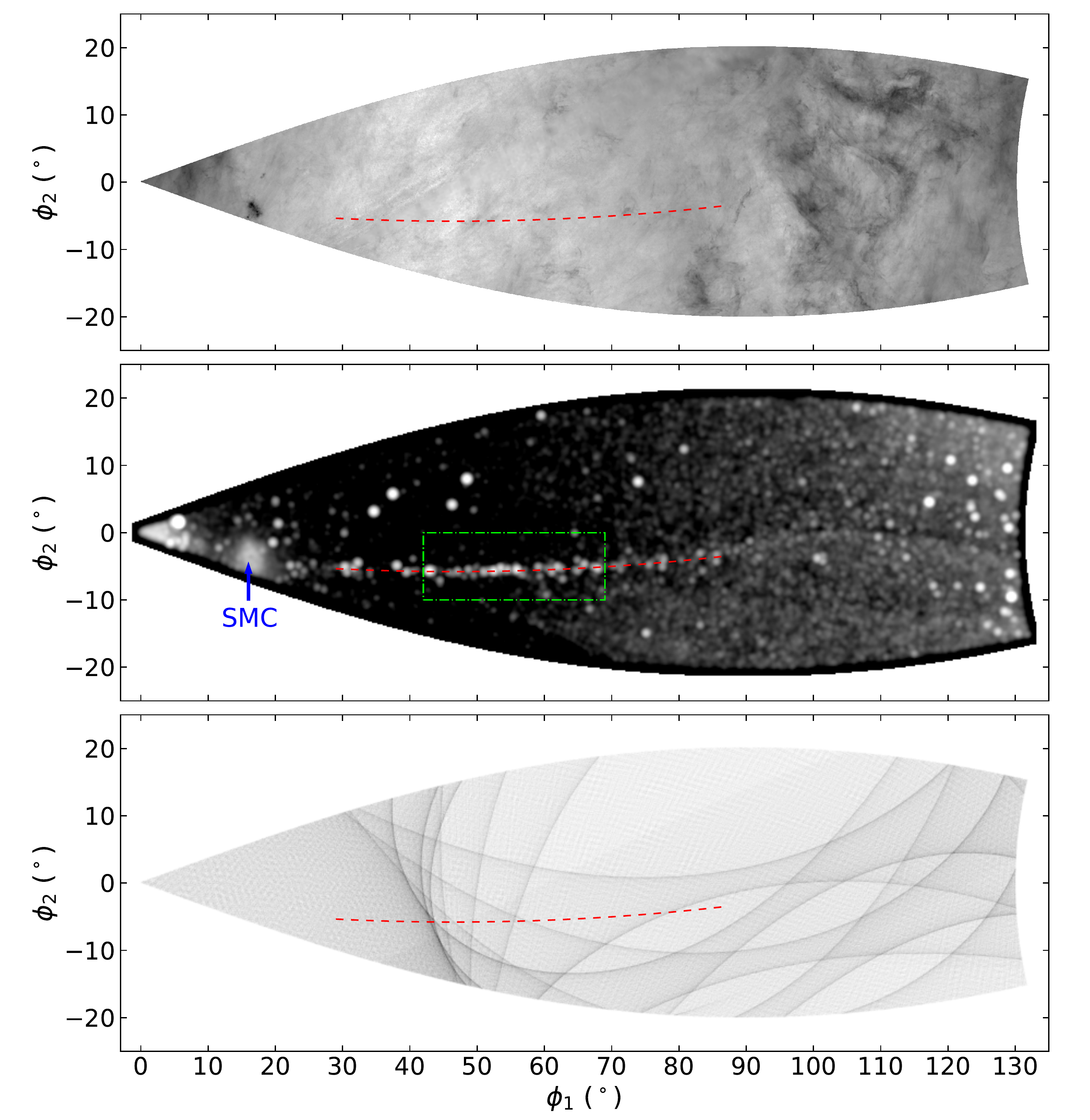}
	\caption{The upper and lower panels present the dust extinction map extracted from \citet{1998ApJ...500..525S} and $Gaia$'s scanning pattern covered by the EDR3, respectively. The middle panel presents the weighted sky map. The blue arrow points the SMC. Stars within the green rectangle are used to create the lateral profile of SGP-S. The red dashed line of the three panels indicates the trajectory of the stream.
		\label{fig:phi1_phi2}}
\end{figure*}

\begin{figure}
	\plotone{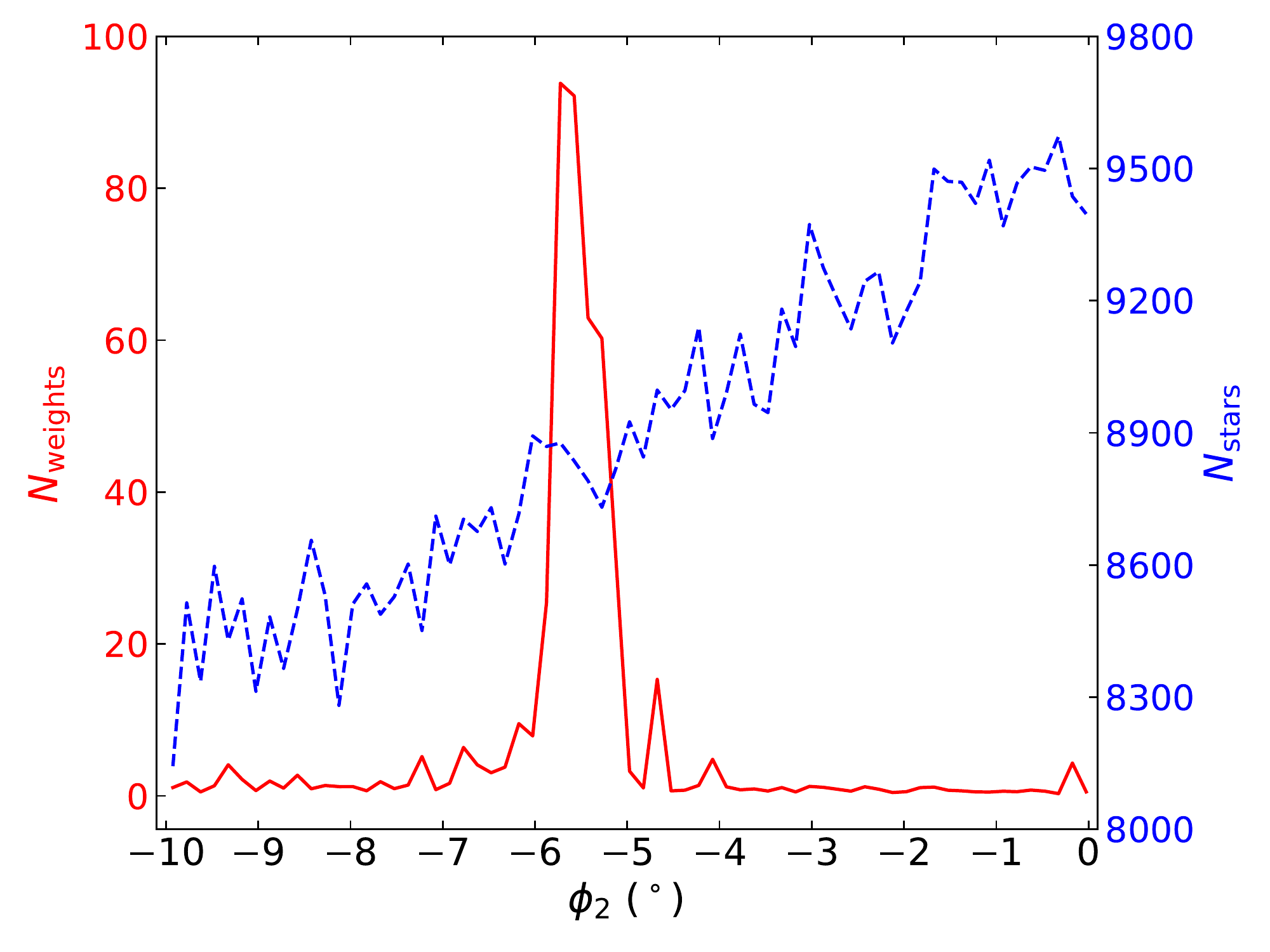}
	\caption{The lateral distributions along $\phi_2$ for weighted (red solid) and unweighted (blue dashed) numbers of stars within the green rectangle of Figure \ref{fig:phi1_phi2}.
		\label{fig:profile}}
\end{figure}

A weighted sky map is obtained as shown in the middle panel of Figure \ref{fig:phi1_phi2}. Here the isochrone filter of Age = 12 Gyr, [M/H] = -2.0 dex and DM = 14.9 mag is used during weighting stars in CMD, which is the best fit result that presents the strongest stream signal\footnote{We measure the stream's strength through the total weights between $-6.2\degr < \phi_2 < -4.8\degr$ in Figure \ref{fig:profile}.}. The sky pixel width is 0.2$\degr$ and the map is smoothed with a Gaussian kernel of $\sigma$ = 0.3$\degr$. The stretch is logarithmic, with brighter areas corresponding to higher weight regions. The blue arrow points the Small Magellanic Cloud (SMC). On the upper and lower panels, we further plot the dust extinction map extracted from \citet{1998ApJ...500..525S} and $Gaia$'s scanning pattern covered by the EDR3, respectively, with higher values represented by darker colors. The red dashed line of the three panels indicates the trajectory of SGP-S fitted by a polynomial of:  
\begin{equation}
	\phi_2 = 1.41834461 \times 10^{-3} \phi_1^2 - 0.132161401 \phi_1 - 2.71971339.
	\label{equation3}
\end{equation}

From the matched-filter map, the signature of SGP-S is quite obvious, starting at $\phi_1 \simeq$ 29$\degr$ and ending at $\phi_1 \simeq$ 87$\degr$. Besides, there are several random noises appearing above and to the right of the stream. They do not represent physical overdensities but are caused by some field stars distributed coincidentally close to our CMD and PM filters, since more stars are populated at higher $\phi_1$ (near the disk) and higher $\phi_2$ (see Figure \ref{fig:profile}) sides. According to the other two panels, we can verify that the stream does not follow any structures in the interstellar extinction and is not aligned with any features in $Gaia$’s scanning pattern. The trajectory in ICRS frame can be well described with a third-order polynomial:
\begin{equation}
	\alpha = - 2.44373567 \times 10^{-5} \delta^3 - 1.62880413 \times 10^{-3} \delta^2 - 0.138532434 \delta + 3.04632126
\end{equation}
where -61$\degr$ $<$ $\delta$ $<$ -3$\degr$. 

To estimate the stream's width, we select stars within a box of 42$\degr$ $<$ $\phi_1$ $<$ 69$\degr$ and -10$\degr$ $<$ $\phi_2$ $<$ 0$\degr$ (green rectangle of Figure \ref{fig:phi1_phi2}), in which the stream is almost parallel to $\phi_1$-axis. We then sum these weights in each $\phi_2$ bin (bin width = 0.15$\degr$) to create a one-dimensional stream profile as shown with the red solid line in Figure \ref{fig:profile}. The stream is almost enclosed within $-6.2\degr < \phi_2 < -4.8\degr$, and its peak is $41\sigma$ above the background noise outside this range. From the profile, we find an estimate of its width (full width at half maximum) to be $\sim 0.6\degr$. 

Furthermore, we create the lateral profile of star's number in the same way and overplot it with the blue dashed line in Figure \ref{fig:profile}. There is a gradient in the distribution of stars along $\phi_2$, with more stars populated at higher $\phi_2$. It can be concluded that the stream signature is not caused by contamination of the Milky Way's field population. Otherwise, it is more likely to detect strong signals close to $\phi_2 = 0\degr$ side. The same analysis is performed for other stream segments and similar profiles for weighted and unweighted numbers of stars can be found. 

\subsection{CMD and PMs}

\begin{figure*}
	\includegraphics[width=\linewidth]{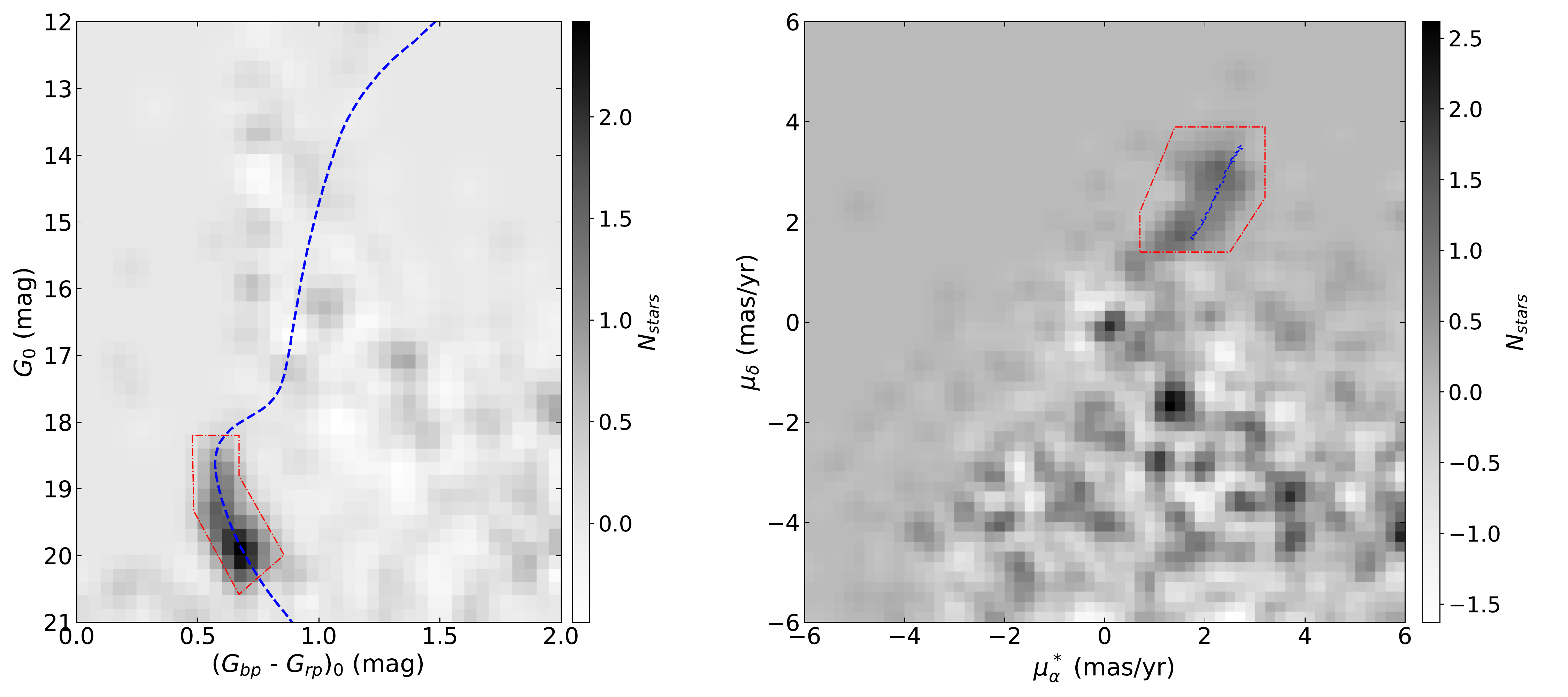}
	\caption{The left panel is a 2D histogram of stars in CMD with PMs selected and background subtracted. The blue dashed line represents the best fit isochrone with Age = 12 Gyr and [M/H] = -2.0 dex at DM = 14.9 mag. The right panel is a 2D histogram of PMs after CMD selection and background subtraction. The filters of Figure \ref{fig:phi1_pm} within $29\degr < \phi_1 < 87\degr$ are overplotted in the blue line. Both of the diagrams are smoothed with a 2D Gaussian kernel of $\sigma$ = 1 pixel. The red polygons represent the CMD and PM selections applied to the stream and off-stream regions.
		\label{fig:CMD_PMs}}
\end{figure*}

We display a background-subtracted binned CMD for the stream in the left panel of Figure \ref{fig:CMD_PMs}. The stream region is defined as the area around the trajectory (Equation \ref{equation3}) $\pm 0.3\degr$ in $\phi_2$, given the derived width of 0.6$\degr$. The background is estimated through averaging two off-stream regions parallel to the stream obtained by moving the stream region along $\phi_2$-axis by $\pm 2\degr$, to eliminate the effect of the gradient. Before the background subtraction, a PM selection is applied to both of the stream and off-stream regions as illustrated with the red polygon in the right panel of Figure \ref{fig:CMD_PMs}, which corresponds to the stream's distribution in PM space (see below). We emphasize that this is a subtraction of star numbers, not weighted counts. 

The CMD bin size is 0.05 mag in color and 0.2 mag in $G$ magnitude. The diagram is smoothed with a 2D Gaussian kernel of $\sigma$ = 1 pixel. The blue dashed line represents the best fit isochrone with Age = 12 Gyr and [M/H] = -2.0 dex at DM = 14.9 mag. After the PM selection and background subtraction, the stream's main sequence along with its turn-off is clearly seen and has a good match with the isochrone. The DM 14.9 mag corresponds to a heliocentric distance of $\sim$ 9.5 kpc. Considering the width of 0.6$\degr$, the physical width of SGP-S is about 100 pc, comparable to dynamically cold streams with GC origins such as Pal 5 \citep[120 pc;][]{2003AJ....126.2385O}, Triangulum \citep[75 pc;][]{2012ApJ...760L...6B}, ATLAS \citep[90 pc;][]{2014MNRAS.442L..85K}, and Molonglo \citep[128 pc;][]{2018ApJ...862..114S}. 

In the right panel of Figure \ref{fig:CMD_PMs}, we present a 2D histogram of PMs. Similarly, before the subtraction between the stream region and the mean of the off-stream regions, a CMD selection is applied to them as shown with the red polygon in the left panel. The diagram with bin size = 0.2 mas/yr is also smoothed using a 2D Gaussian with $\sigma$ = 1 pixel. The filters of Figure \ref{fig:phi1_pm} within $29\degr < \phi_1 < 87\degr$ are overplotted in the blue dashed line. An overdensity at $\mu^*_{\alpha} \sim$ 2.0 mas/yr and $\mu_{\delta} \sim$ 2.5 mas/yr is discernable corresponding to the stream. We note that our PM filters happen to be a rough estimate of the stream's PMs. It is worth noting that the PM filters are necessary for exposure of SGP-S because the signals are indistinguishable when using the CMD filter alone, and the PM filters assign rather low weights to most of field stars so that the stream can be discerned. 

The stream's surface density and brightness are further estimated. There are a total of 106 stars within the PM polygon after the background subtraction. This serves as an estimate to the number of the stream stars located in a $58\degr \times 0.6\degr$ region observed by $Gaia$. Thus the surface density is roughly 3 stars degree$^{-2}$. For all stars in the stream region, each one is assigned a weight by the matched-filter method and this allows us to select the most likely members of the stream based on the sort of weights. We adopt stars with weights $>$ 0.08 as the member candidates because the criterion leaves us 106 stars as well. By combining their individual $G$ magnitudes, we get a surface brightness of SGP-S to be $\Sigma_G \simeq$ 36.2 mag arcsec$^{-2}$, which is even darker than, for example, Phlegethon \citep[34.3 mag arcsec$^{-2}$;][]{2018ApJ...865...85I} and the trailing tail of M5 \citep[35 mag arcsec$^{-2}$;][]{2019ApJ...884..174G}.

\subsection{Association with GCs and Streams}

We aim to fit an orbit to SGP-S so that we can investigate whether it is related to any GCs or known streams of the Milky Way. We assume a Galactic potential model of \texttt{MWPotential2014} \citep{2015ApJS..216...29B}. The solar distance to the Galactic centre, circular velocity at the Sun, and solar velocities relative to the local standard of rest are set to 8 kpc, 220 km s$^{-1}$ \citep{2012ApJ...759..131B}, and (11.1, 12.24, and 7.25) km s$^{-1}$ \citep{2010MNRAS.403.1829S}, respectively. The fitting parameters are position $\alpha$, $\delta$, heliocentric distance $d$, PMs $\mu^*_{\alpha}$, $\mu_{\delta}$, and radial velocity $V_r$. We chose to anchor the declination at $\delta$ = -61$\degr$, an endpoint of the stream, leaving other parameters free to be varied. In a Bayesian framework, sky positions and PMs of 106 member candidates are used to constrain the parameters and the fitted results can be derived from their marginalized posterior distributions through a Markov Chain Monte Carlo sampling. The best-fit parameters are $\alpha = 9.63^{+0.09}_{-0.09}$ $\degr$, $d = 10.17^{+0.12}_{-0.12}$ kpc, $\mu^*_{\alpha} = 2.47^{+0.04}_{-0.04}$ mas yr$^{-1}$, $\mu_{\delta} = 3.53^{+0.07}_{-0.07}$ mas yr$^{-1}$ and $V_r = 34.06^{+3.64}_{-3.68}$ km s$^{-1}$. 

\begin{figure*}
	\includegraphics[width=\linewidth]{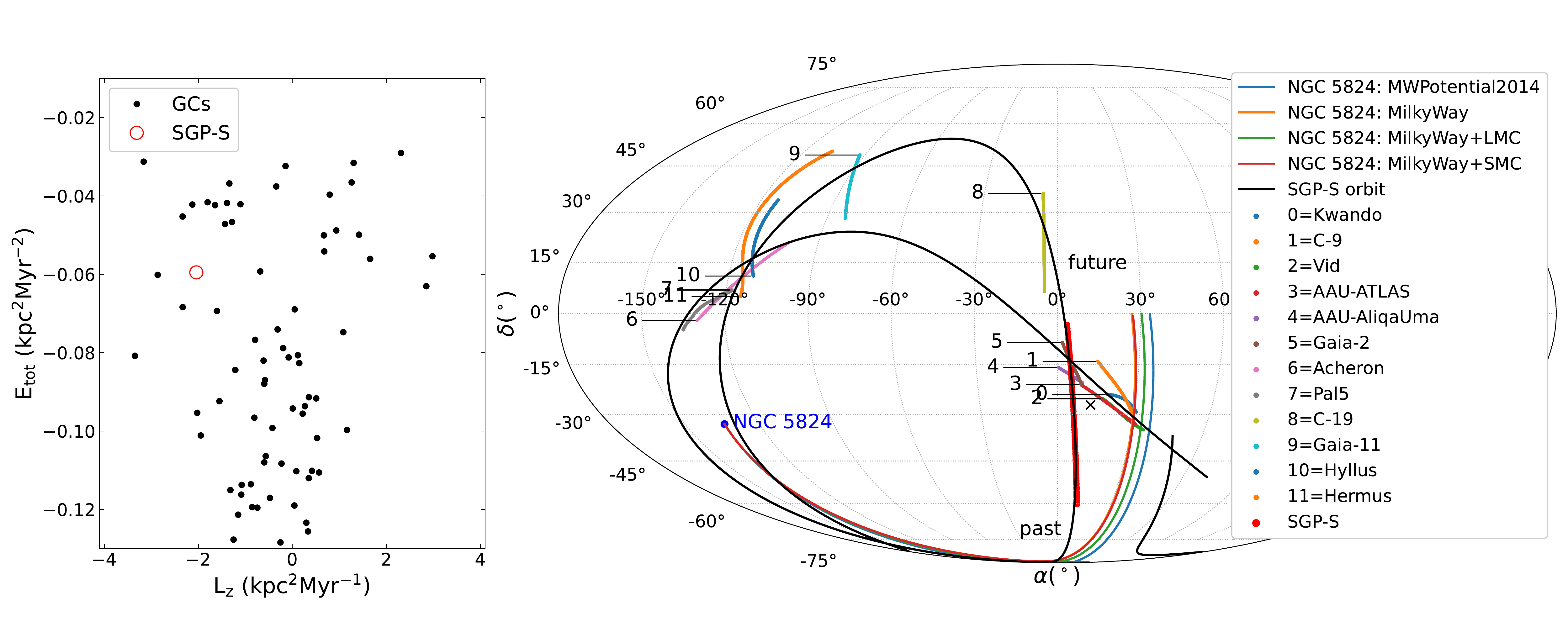}
	\caption{The left panel displays GCs (black points) and SGP-S (red circle) in angular momenta and energy space. The right panel presents projections of streams and orbits in ICRS frame. The red track and black line indicate SGP-S and its best-fit orbit. Colored lines represent orbital paths for GC NGC 5824 integrated in different potential models. Numbered tracks represent 12 known streams obtained from \citet{2022arXiv220410326M}. The black cross denotes the southern Galactic pole. 
		\label{fig:GCs_streams}}
\end{figure*}

We first examine possible connections between SGP-S and GCs by comparing their angular momenta $L_z$ and energy $E_{\rm tot}$ as shown in the left panel of Figure \ref{fig:GCs_streams}. The positions and velocities of GCs are taken from \citet{2021MNRAS.505.5978V}. It can be seen that the stream does not lie close to any other GCs. We have further integrated orbits of all 160 GCs and then compared them to the trajectory of SGP-S but no consistent orbits are found. We consider that the progenitor for SGP-S might have been dissolved. 

In the right panel, we present the trajectory of SGP-S (red track) along with its best-fit orbit (black line) integrated for $\pm$1 Gyr to the past and future. Its pericenter and apocenter are $R_{peri}$ = 10.5 kpc and $R_{apo}$ = 49.6 kpc, respectively. Since the detection is motivated by searching for GC NGC 5824's tidal debris, its orbital path is also overplotted for a comparison. We explore the cluster's orbits using 4 potential models. The first one is \texttt{MWPotential2014} used in this work (blue line). The second one come from the static Milky Way potential as applied in \citet{2022MNRAS.513..853Y} which we refer to as \texttt{MilkyWay} (orange line). The other two potentials further contain the LMC (green line) or SMC (red line) components on the basis of \texttt{MilkyWay}, both of which are modelled using a \citet{1990ApJ...356..359H} sphere with masses and scale sizes from \citet{2022MNRAS.510.2437E}. Apparently, although SGP-S is found coincidentally using the PMs of NGC 5824 model stream, they are completely separated on sky, even if variations and perturbations in potential models are taken into consideration. 

Furthermore, 12 known streams nearly aligned with SGP-S's orbit on sky are also presented as marked in different numbers, which are obtained from \citet[][and references therein]{2022arXiv220410326M}. Although some stream projections seem to fit the orbit well, they are actually not connected when considering additional criteria. Specifically, $R_{peri}$ and $R_{apo}$ values for Kwando (4.4, 26.4), C-19 (7.0, 27.4), Hermus (7.1, 17.2) and Hyllus (5.4, 18.6) kpc are inconsistent with those of SGP-S. The other streams are separated from the orbit in heliocentric distance: C-9 (8.1 versus 15.2)\footnote{The median distance of streams versus the orbital distance of SGP-S.}, Vid (24.5 versus 16.1), ATLAS (20.2 versus 15.3), Aliqa Uma (26.6 versus 12.1), Gaia-2 (7.0 versus 12.0), Acheron (3.6 versus 33.5), Pal 5 (21.2 versus 43.1), and Gaia-11 (12.4 versus 36.0) kpc. Hence, it is concluded that there is no favorable match to newly discovered SGP-S among known streams. 


\section{Conclusion} \label{sec:summary}

With revised photometry and astrometry from $Gaia$ EDR3, we have discovered a new cold stream near the southern Galactic pole which we dub SGP-S. The stream is detected at a significance of 41$\sigma$ through a modified matched-filter that assigns weights to stars in CMD and PMs simultaneously. The SGP-S is spanning 58$\degr$ by 0.6$\degr$ on sky at 9.5 kpc away from the sun. The best-fit isochrone indicates an old ($\sim$ 12 Gyr) and metal-poor ($\sim$ -2.0 dex) stellar population. The stream has an extremely low surface brightness of $\Sigma_G \simeq$ 36.2 mag arcsec$^{-2}$, with a density of about 3 stars degree$^{-2}$. Given the physical width to be 100 pc, we further explore the possibility of connections between SGP-S and extant GCs along with other narrow streams of the Milky Way and find that none of them has similar dynamical properties to the stream. Follow-up observations and studies might be able to uncover the origin of SGP-S.

\begin{acknowledgments}
We thank the referee for the thorough reviews that helped us to improve the manuscript. This study is supported by the National Natural Science Foundation of China under grant nos 11988101, 11973048, 11927804, 11890694 and 11873052, and the National Key R\&D Program of China, grant no. 2019YFA0405500. This work is also supported by the GHfund A (202202018107). We acknowledge the support from the 2m Chinese Space Station Telescope project CMS-CSST-2021-B05. 

This work presents results from the European Space Agency (ESA) space mission Gaia. Gaia data are being processed by the Gaia Data Processing and Analysis Consortium (DPAC). Funding for the DPAC is provided by national institutions, in particular the institutions participating in the Gaia MultiLateral Agreement (MLA). The Gaia mission website is \url{https://www.cosmos.esa.int/gaia}. The Gaia archive website is \url{https://archives.esac.esa.int/gaia}.

\end{acknowledgments}

%




\bibliography{sample631}{}
\bibliographystyle{aasjournal}



\end{document}